\documentclass[%
preprint,
 amsmath,amssymb,
 aps,
]{revtex4-2}

\usepackage{stackengine}
\usepackage{graphicx}
\usepackage{dcolumn}
\usepackage{bm}

\begin{document}

\title{Time modulated media with digital meta--atoms}

\author{S. A. R. Horsley}
\email{s.horsley@exeter.ac.uk}
\affiliation{School of Physics and Astronomy, University of Exeter, Stocker Road, EX4 4QL}

\begin{abstract}
We develop the theory of acoustic wave propagation in a waveguide containing an array of time modulated digital meta--atoms, showing the equivalence between this array and a homogeneous, time varying, dispersive material.  In the limit of an adiabatic time variation we find some choices of meta-atom coupling strength lead to exceptional points, where one mode is exponentially amplified and the adiabatic approximation breaks down.  In the highly non--adiabatic limit we derive the analogue of reflection coefficients for an abrupt change in the meta-atom coupling.  Due to dispersion we have both conversion between different \emph{types} of modes, and different \emph{directions} of propagation, which can be tuned through modifying the programmed response of the digital meta--atoms.
\end{abstract}

\maketitle

%
%
\section{Introduction}
A material that changes in space, where the wave speed is non--uniform, can act as a lens, mirror, absorber, or invisibility cloak, depending on the type of material and how it is arranged.  Such structures can be traced back to ancient investigations of `burning lenses'~\cite{smith2017} and remains the subject of most research into electromagnetic (EM) and acoustic materials~\cite{kadic2019}.  Yet until recently, the variation of material properties in \emph{time} had received less attention.

Morgenthaler~\cite{morgenthaler1958} established that homogeneous time varying media conserve electromagnetic momentum, rather than energy.  As a time variation of the dielectric function appears naturally in the properties of an unsteady plasma, this early work was first applied in plasma physics (e.g.~\cite{stepanov1976}), findings that are generalized and summarized in the concepts of ``time refraction'' and ``time reflection'' described in the work of Mendon\c{c}a and others~\cite{mendonca2000,mendonca2002}.  Here a temporal jump in material parameters generates additional waves, analogous to scattering from a spatial boundary, but rotated from space into time and subject to the constraints of causality.  Connected to this work, the field of analogue gravity has found deep results concerning materials with a space--time varying wave speed~\cite{schutzhold2005,philbin2008} as a means to mimic a moving medium with an associated analogue event horizon and Hawking radiation~\cite{weinfurtner2011,drori2019}.

In recent years there has been renewed interest in time varying media from those working on artificial electromagnetic and acoustic materials~\cite{xiao2014,ptitcyn2019}.  It has been quikcly established that this enables new routes to thin absorbers~\cite{li2021}, gain without a dispersive response~\cite{pendry2021}, the phenomenon of `temporal aiming'~\cite{pena2020a}, temporal anti--reflection coatings~\cite{pena2020b}, and a homogenized response equivalent to bianisotropy~\cite{huidobro2021}.  It is now understood that besides being an interesting variant of the usual reflection problem, rapid temporal changes in material properties modify EM and acoustic fields in a distinct way from spatial variations.

We can understand the difference between material variation in space versus time from both Maxwell's equations
\begin{equation}
    \left(\begin{matrix}\boldsymbol{0}&\boldsymbol{\nabla}\times\\-\boldsymbol{\nabla}\times&\boldsymbol{0}\end{matrix}\right)\left(\begin{matrix}\boldsymbol{E}\\\boldsymbol{H}\end{matrix}\right)=\frac{\partial}{\partial t}\left(\begin{matrix}\boldsymbol{D}\\\boldsymbol{B}\end{matrix}\right)\label{eq:maxwell}
\end{equation}
and the equations of elasticity
\begin{equation}
    \frac{\partial}{\partial t}\left(\rho\frac{\partial\boldsymbol{U}}{\partial t}\right)=\boldsymbol{\nabla}\cdot\boldsymbol{\sigma}\label{eq:elasticity}.
\end{equation}
The \emph{spatial} derivatives in Eq. (\ref{eq:maxwell}) remain finite so long as the tangential electric and magnetic fields are continuous in space, and likewise for the normal stresses in Eq. (\ref{eq:elasticity}).  This observation leads to the usual boundary conditions of elasticity and electromagnetism across interfaces between different media, and the continuity of power flow across any interface.  Conversely the \emph{temporal} derivatives remain finite so long as the $\boldsymbol{D}$ and $\boldsymbol{B}$ vectors, or equivalently momentum density $\rho\dot{\boldsymbol{U}}$ are continuous in time~\cite{xiao2014,bakunov2014}.  These are different boundary conditions, and do not lead to e.g. continuity of power flow in time.  In the electromagnetic case these boundary conditions ensure the Abraham momentum, $\boldsymbol{E}\times\boldsymbol{H}$ is conserved across a spatial boundary, while the Minkowskii momentum, $\boldsymbol{D}\times\boldsymbol{B}$ is conserved across a temporal one~\cite{griffiths2012}.  This potential to change power flow underlies the phenomenon of `temporal aiming'~\cite{pena2020a}, where a temporal change in the principal axes of the material tensor can be used to redirect the flow of wave energy.  Similarly---unlike their spatial counterpart---the transmission through a temporal anti--reflection coefficient is not unity~\cite{mai2021}, indicating a change in the power flow due to the material time variation.

\begin{figure}[h!]
    \centering
    \includegraphics[width=\textwidth]{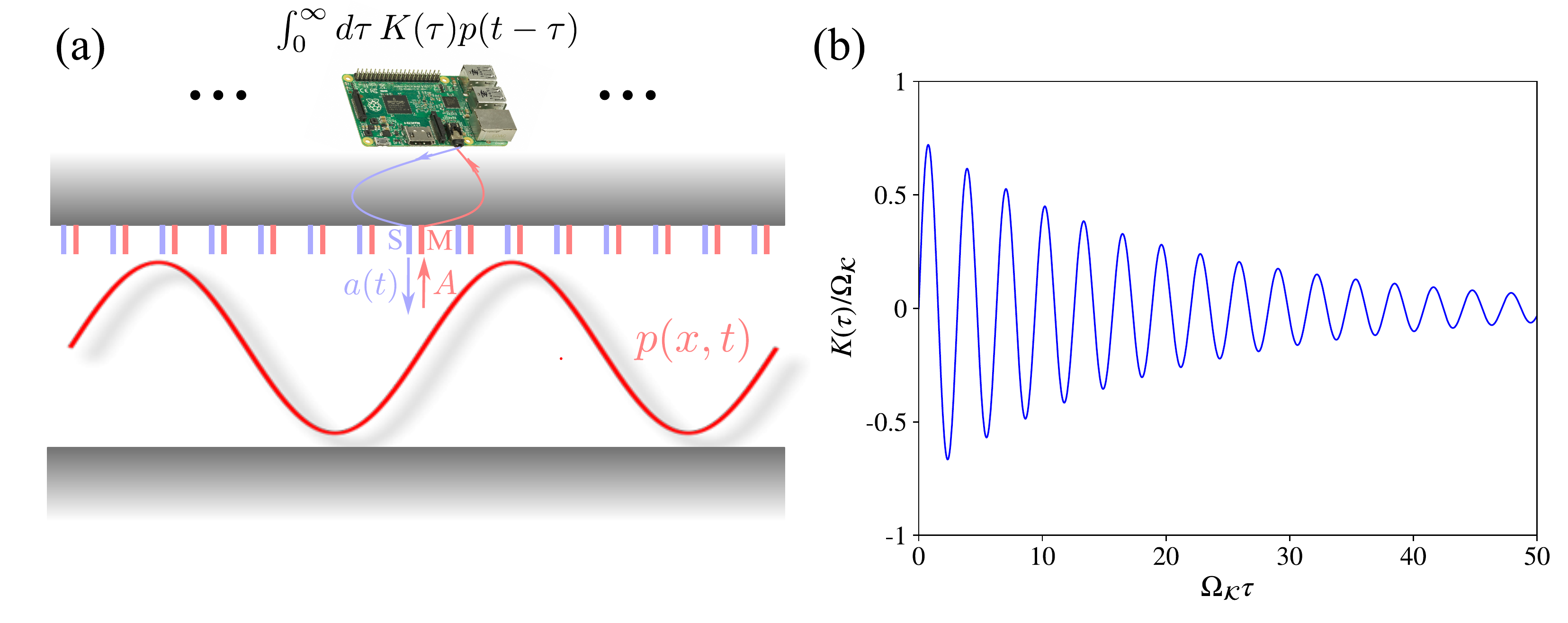}
    \caption{A waveguide containing an array of digital meta--atoms, each with response kernel $K(\tau)$. (a) Schematic: an acoustic waveguide contains an array of speakers $S$ and microphones $M$, each connected by a single board computer, implementing response kernel $\mathcal{K}(\tau)$, and modulated speaker amplitude $a(t)$.  (b) The signal sent to the speaker is equal to the past values of the pressure field $p(t)$ measured by the microphone, convolved with the response kernel $K(t)$, which is here a damped sinusoid (a Lorentzian in the frequency domain).  We plot the function defined in Eq. (\ref{eq:lorentz_kernel}) with parameters   $\omega_0/\Omega_{\mathcal{K}}=2.0$, $\gamma/\Omega_{\mathcal{K}}=0.1$,  and $A/\Omega_{\mathcal{K}}^2=1.5$.}
    \label{fig:schematic}
\end{figure}

To date there are only a few experimental realisations of time dependent metamaterials.  In electromagnetism these typically use on a non--linear response~\cite{zhou2020} to induce the time variation.  In acoustics---where the frequencies are usually reduced by many orders of magnitude---there is more freedom to control the time variation of the material.  We base the theory in this paper on the concept of active acoustic meta--atoms~\cite{popa2013}, recently implemented by the group of Li and co--workers~\cite{cho2020,wen2020,wen2021} to demonstrate metamaterials with a non--Hermitian and time varying response.  Each `digital meta--atom' consists of two closely spaced speaker--microphone pairs inserted into an acoustic waveguide, connected via a single--board computer.  The two speakers measure the acoustic pressure field at two points, computing its average and gradient.  From this the single board computer computes a response as a function of the measured pressure field, and sends signals to the microphone pair.  The response can be programmed with great freedom, convolving the input signal with a pre--specified kernel (e.g. a damped sinusoid as shown in Fig.~\ref{fig:schematic}b) and simultaneously modulating the amplitudes of the speaker output.  An arbitrary dispersive, time modulated response can thereby be obtained at kHz frequencies.  Such systems are similar to those commonly used in active noise and vibration control~\cite{hansen2012}, but here applied to realise a tailored time dependent dispersive response.  Based on this idea we give a simple theoretical model for an array of such scatterers with time modulated properties, as described in Figure~\ref{fig:schematic}.

%
%
%
%
\section{Digital acoustic meta--atoms}

We first construct a theoretical model of a single digital meta--atom in a waveguide.  Consider acoustic pressure waves, obeying the  linearized Navier--Stokes equations for a fluid of uniform density $\rho_0$, pressure $p_0$, and zero local flow velocity.  For small deviations around this background, the Navier--Stokes equation reduces to a linear equation for the fluid velocity and pressure
\begin{equation}
    \rho_0\partial_t\boldsymbol{v}=-\boldsymbol{\nabla}p+\boldsymbol{f}\label{eq:linear_ns}.
\end{equation}
where $\boldsymbol{f}$ is the external force density applied to the fluid.  In this linear limit the continuity equation becomes
\begin{equation}
    \partial_t\rho+\rho_0\boldsymbol{\nabla}\cdot\boldsymbol{v}=-q\label{eq:linear_c}.
\end{equation}
where $q$ represents a source of mass due to e.g. the fluid moving in and out of a compressed region.  Assuming one dimensional propagation (i.e. the fundamental mode in a waveguide), writing $\rho=(\partial\rho/\partial p)_S\, p$, and combining (\ref{eq:linear_ns}--\ref{eq:linear_c}) we find the wave equation for the pressure field,
\begin{equation}
    \left(\partial_x^2-\frac{1}{c_{s}^2}\partial_t^2\right) p=\partial_t q+\partial_x f_x\label{eq:driven_wave}
\end{equation}
where the speed of sound is given by $c_s=\left(\partial p/\partial \rho\right)_{S}^{1/2}$.  Here the source of the sound is the forcing due to a speaker at $x=0$, driven by a signal $\mathcal{F}(t)$.  We take this as a point--like monopole excitation, $\partial_x f_x=0$ and
\begin{equation}
    \partial_t\,q=\delta(x)\mathcal{F}(t)\label{eq:forcing}.
\end{equation}
Assuming an incoming plane wave $p_{\rm in}=\cos(\omega(x/c_s-t))$, plus the wave generated from the forcing (\ref{eq:forcing}), the solution to Eq. (\ref{eq:driven_wave}) is
\begin{equation}
    p(x,t)=\cos(\omega(x/c_s-t))-\frac{c_s}{2}\int_{0}^{\infty}\mathcal{F}(t-|x|/c_s-\tau)\,d\tau\label{eq:pressure_field}
\end{equation}
Up until this point we have not considered the dependence of the speaker signal $\mathcal{F}(t)$ on the incident pressure field.  Via the signal processing on the single--board computer, the speaker amplitude is related linearly to the past behaviour to the pressure field by the following integral equation
\begin{equation}
    \mathcal{F}(t)=\frac{a(t)}{w}\partial_t^2\int_{0}^{\infty}d\tau\,K(\tau)\,p(0,t-\tau)\label{eq:constitutive_relation}
\end{equation}
where $K(\tau)$ has dimensions of frequency and represents the causal response of the digital meta--atom.  The causal response can be chosen almost arbitrarily through programming the single--board computer.  As $\mathcal{F}$ has dimensions of force per volume, we have introduced a characteristic length $w$ of the digital meta--atom.  The dimensionless function $a(t)$ is an additional applied time modulation of the signal sent to the speaker pair, and allows the meta--atom to exhibit a pre--specified time dependent response.
%
%
\section{Effective medium limit of a digital meta--atom array}
\par
We take a one--dimensional array of these meta--atoms spaced by distance $\Delta$, with one speaker--microphone pair per unit cell.  We shall show that in the long wavelength limit the scatterers are equivalent to a homogeneous medium with time varying properties.  Inserting Eq. (\ref{eq:constitutive_relation}) into (\ref{eq:driven_wave}--\ref{eq:forcing}) and summing the force over the array of speakers we find the pressure field in the array obeys,
\begin{equation}
    \left(\partial_{x}^2-\frac{1}{c_s^2}\partial_t^2\right)p=\frac{a(t)}{w}\sum_{n=-\infty}^{\infty}\delta(x-n\Delta)\,\partial_t^2\int_{0}^{\infty}d\tau\,K(\tau)\,p(n\Delta,t-\tau)\label{eq:scatterer-array}
\end{equation}
Defining $\bar{p}$ as the pressure averaged over a single unit cell,
\begin{equation}
    \bar{p}(x,t)=\frac{1}{\Delta}\int_{-\Delta/2}^{\Delta/2}p(x+y,t)\,dy
\end{equation}
we find that it obeys
\begin{equation}
    \left(\partial_x^2-\frac{1}{c_s^2}\partial_t^2\right)\bar{p}(x,t)=\frac{a(t)}{w \Delta}\partial_t^2\int_0^{\infty}d\tau\,K(\tau)\,p([x/\Delta]\Delta,t-\tau)\label{eq:average_pressure}
\end{equation}
where $[x/\Delta]$ is the nearest integer to the real number $x/\Delta$.  From Eq. (\ref{eq:pressure_field}) we can see that the change in the pressure field over a unit cell is determined by the time variation of each speaker output over the interval $\Delta/c_s$.  When neither the speaker amplitude, $a(t)$ or the pressure, $p(t)$ vary significantly over this timescale we can replace $p([x/\Delta]\Delta,t-\tau)$ with $\bar{p}(x,t-\tau)$ and the average pressure obeys
\begin{equation}
    \left(\partial_x^2-\frac{1}{c_s^2}\partial_t^2\right)\bar{p}(x,t)=\frac{\bar{a}(t)}{c_s^2}\partial_t^2\int_0^{\infty}d\tau\,K(\tau)\,\bar{p}(x,t-\tau)\label{eq:effective_medium}.
\end{equation}
where $\bar{a}=(c_s^2/w\Delta)\,a$.  This wave equation is equivalent to that in a homogeneous dispersive medium (dispersion determined by the Fourier transform of $K(\tau)$), which is modulated in time according to $a(t)$.  For memory kernels $K(\tau)$ with a short time response, the right hand side of (\ref{eq:effective_medium}) can be expanded as a series, $\int_0^\infty 
d\tau\,K(\tau)u_{\mathcal{K}}(t-\tau)=\sum_{n}(-1)^n\partial_{t}^{n}u_{\mathcal{K}}(t)\int_0^\infty 
d\tau\,K(\tau)\tau^n/n!$.  Taking just the first term in this series reduces Eq. (\ref{eq:effective_medium}) to that for a scalar wave in a medium with a non--dispersive but time modulated refractive index, as described in e.g.~\cite{holberg1966}.  Taking the next two terms in the series introduces successive dispersive terms represented as third and fourth order time derivatives of the pressure in Eq. (\ref{eq:effective_medium}).

As in the work of Cho et al.~\cite{cho2020}, we take our kernel to be of Lorentzian form in the frequency domain, which corresponds to a damped sinusoidal time response
\begin{equation}
    K(\tau)=A\Theta(\tau)\frac{\sin(\omega_1\tau)}{\omega_1}{\rm e}^{-\frac{\gamma}{2} \tau}\label{eq:lorentz_kernel}
\end{equation}
where $\omega_1=\sqrt{\omega_0^2-(\gamma/2)^2}$, and $A$ is a constant with dimensions of frequency squared, setting the scale of the response of each digital meta--atom.  Writing the pressure field as $\bar{p}=u_{\mathcal{K}}(t){\rm e}^{{\rm i}\mathcal{K} x}$, where $\mathcal{K}\Delta\ll 1$ and defining the auxiliary quantity $v(t)=\int_0^{\infty}d\tau\,K(\tau)\,\partial_t^2 u_{\mathcal{K}}(t-\tau)$ we find that (\ref{eq:effective_medium}) is now equivalent to a non--Hermitian system of two coupled simple harmonic oscillators
\begin{align}
    \left(\partial_t^2+\Omega_{\mathcal{K}}^2\right)u_{\mathcal{K}}(t)&=-\bar{a}(t) v(t)\nonumber\\
    \left(\partial_t^2+\gamma\partial_t+\omega_0^2\right)v(t)&=A \partial_t^2 u_{\mathcal{K}}(t)\label{eq:two_oscillator_system}
\end{align}
where $\Omega_{\mathcal{K}}=c_s\mathcal{K}$.  While the oscillator $u_{\mathcal{K}}(t)$ represents the amplitude of the sound wave within the lattice, propagating with wave--vector $\mathcal{K}$, the amplitude $v(t)$ mimics the dynamics of the speaker and microphone system, coupled through the single--board computer array.
%
%
\section{Adiabatic modulation}
%
%
\begin{figure}[h!]
    \centering
    \includegraphics[width=\textwidth]{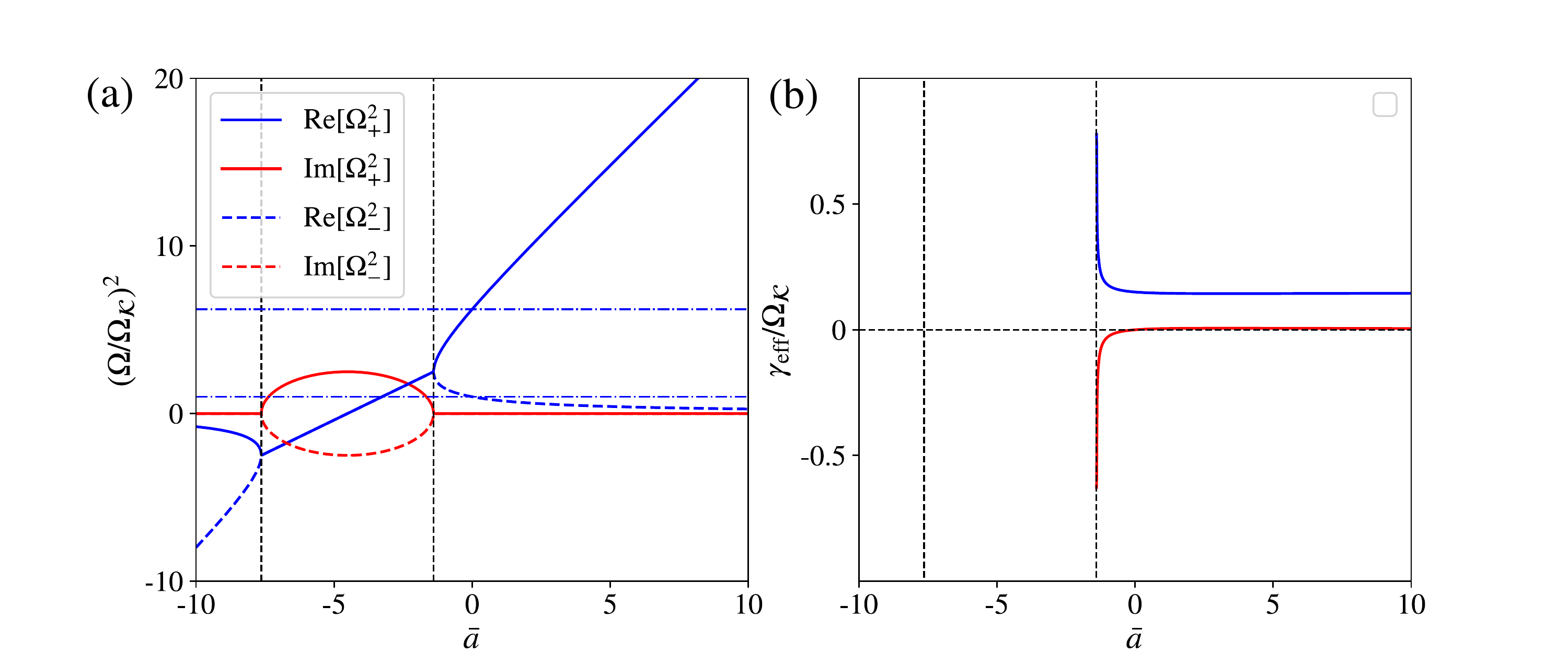}
    \caption{Coupled mode frequencies $\Omega_{\pm}^2$ and effective damping $\gamma_{\rm eff}$ for parameters $A=1.6\Omega_{\mathcal{K}}^2$, $\omega_0=2.5\Omega_{\mathcal{K}}$, and $\gamma=0.3\Omega_{\mathcal{K}}$. Panel (a) shows real and imaginary parts of the coupled mode frequencies computed from (\ref{eq:Wpm}), illustrating the two roots for choice of speaker output $\bar{a}$.  Vertical dashed lines show the two exceptional point values of $\bar{a}$ computed from Eq. (\ref{eq:aex}).  Horizonal dashed lines indicate the uncoupled oscillator frequencies $\Omega=\Omega_{\mathcal{K}}$ and $\Omega=\sqrt{\omega_0^2-\gamma^2/4}$.  Panel (b) shows the effective damping coefficient $\gamma_{\rm eff}$, calculated using (\ref{eq:gamma_eff}), for the pairs of modes shown in (a).  This quantity diverges as we approach an exceptional point, and then loses its meaning.  We have not plotted values in this region.}
    \label{fig:dispersion}
\end{figure}
\par
To gain some insight into the system of equations (\ref{eq:two_oscillator_system}) for the pressure amplitude $u_{\mathcal{K}}$, and auxiliary variable $v$ we take the simplest case where the speaker amplitude $\bar{a}$ is modulated slowly in time.  Our approximation is analogous to the WKB approximation~\cite{heading2013,bender1999} often used in optics and quantum mechanics.  We first write Eq. (\ref{eq:two_oscillator_system}) in matrix form, defining the vector $|\psi\rangle=(u_{\mathcal{K}},v)^{\rm T}$
\begin{equation}
    \left(\begin{matrix}1&0\\-A&1\end{matrix}\right)\partial_{t}^{2}|\psi\rangle+\left(\begin{matrix}0&0\\0&\gamma\end{matrix}\right)\partial_t|\psi\rangle+\left(\begin{matrix}\Omega_{\mathcal{K}}^2&\bar{a}\\0&\omega_0^2\end{matrix}\right)|\psi\rangle=0.\label{eq:matrix_form}
\end{equation}
Inverting the constant matrix in front of the second time derivative in Eq. (\ref{eq:matrix_form}) and performing the transformation $|\psi\rangle=\exp\left(-\frac{1}{2}\Gamma\,t\right)|\psi'\rangle$, where $\Gamma={\rm diag}(0,\gamma)$, eliminates the first order derivative from (\ref{eq:matrix_form}), leaving the second order matrix equation
\begin{equation}
    \frac{d^2}{dt^2}|\psi'\rangle+M(t)|\psi'\rangle=0\label{eq:eqm_new_form}
\end{equation}
where the time dependent $2\times2$ matrix $M(t)$ is given by
\begin{equation}
    M(t)=\left(\begin{matrix}\Omega_{\mathcal{K}}^2&\bar{a}(t){\rm e}^{-\frac{\gamma\,t}{2}}\\A\Omega_{\mathcal{K}}^2{\rm e}^{\frac{\gamma t}{2}}&\omega_0^2-\frac{1}{4}\gamma^2+A\bar{a}(t)\end{matrix}\right).\label{eq:Mt}
\end{equation}
Note that due to the coupling between the oscillators, this matrix is time dependent even when the coupling functions $A$ and $\bar{a}$ are constant in time.  The treatment of this section is thus restricted to small values of the damping $\gamma$

To solve Eq. (\ref{eq:eqm_new_form}) we expand the vector $|\psi'\rangle$ in terms of the instantaneous eigenvectors $|u_{\pm}\rangle$ of the matrix $M(t)$, obeying $M|u_{\pm}\rangle=\Omega_{\pm}^2|u_{\pm}\rangle$.  These eigenvectors are~\footnote{As defined here, when the modulation function $\bar{a}$ passes through zero, one of the eigenvectors $|u_{\pm}\rangle$ defined in (\ref{eq:ev}) itself passes through zero, leading to a discontinuity in the overall phase of this eigenvector.  To deal with this we add a small imaginary part $\bar{a}\to\bar{a}+{\rm i}\eta$, and take the limit $\eta\to0$.}
\begin{equation}
    |u_{\pm}(t)\rangle=\left(\begin{matrix}\bar{a}(t){\rm e}^{-\frac{\gamma t}{2}}\\\Omega^2_{\pm}(t)-\Omega_{\mathcal{K}}^2\end{matrix}\right),\label{eq:ev}
\end{equation}
with associated eigenvalues $\Omega^2_{\pm}$ given by
\begin{equation}
    \Omega_{\pm}^2(t)=\frac{1}{2}\left[\Omega_{\mathcal{K}}^2+\omega_0^2-\frac{1}{4}\gamma^2+A\bar{a}\pm\sqrt{\left(\Omega_{\mathcal{K}}^2+\omega_0^2-\frac{1}{4}\gamma^2+A\bar{a}\right)^2-4\Omega_{\mathcal{K}}^2\left(\omega_0^2-\frac{1}{4}\gamma^2\right)}\right].\label{eq:Wpm}
\end{equation}
The two values $\Omega_{\pm}^2$ are plotted in Fig.~\ref{fig:dispersion}a.  Due to the dispersive response of the meta--atoms, there are \emph{four} possible frequencies of oscillation, arising from the two square roots of (\ref{eq:Wpm}).  These four roots represent the two directions of propagation for each value of $\mathcal{K}$, for both the upper and lower branches of the dispersion relation shown in Fig.~\ref{fig:dispersion}.   Due to the non--Hermitian equation of motion (\ref{eq:two_oscillator_system}) there are exceptional point degeneracies in (\ref{eq:Wpm}) for some choices of speaker amplitude $\bar{a}$
\begin{equation}
    \bar{a}_{\rm ex}=-\frac{1}{A}\left(\Omega_{\mathcal{K}}\pm\sqrt{\omega_0^2-\frac{1}{4}\gamma^2}\right)^2.\label{eq:aex}
\end{equation}
at which point the two eigenvectors (\ref{eq:ev}) become parallel.  Note that these exceptional points require a negative value of the real coupling constant $\bar{a}_{\rm ex}$.  These two exceptional point degeneracies are indicated with the vertical dashed lines in Fig.~\ref{fig:dispersion}a.  
\par
To derive the form of the wave during an adiabatic modulation of $\bar{a}(t)$ we write the vector $|\psi'\rangle$ as a superposition of the instantaneous eigenvectors (\ref{eq:ev})
\begin{equation}
    |\psi'\rangle=c_{+}(t)\frac{{\rm e}^{-{\rm i}\int_0^t\Omega_{+}dt'}}{\sqrt{\Omega_{+}}}|u_+\rangle+c_{-}(t)\frac{{\rm e}^{-{\rm i}\int_0^t\Omega_{-}dt'}}{\sqrt{\Omega_{-}}}|u_-\rangle,\label{eq:adiabatic_form}
\end{equation}
Making the substitution (\ref{eq:adiabatic_form}) in Eq. (\ref{eq:eqm_new_form}) and dropping terms that are second order in the rate of change of $|u_{\pm}\rangle$, we find a first order equation for the amplitudes $c_{\pm}$, with the time variation proportional to the rate of change of the vectors $|u_{\pm}\rangle$
\begin{equation}
    \frac{d}{dt}\left(\begin{matrix}c_{+}\\c_{-}\end{matrix}\right)=-\left(\begin{matrix}\langle v_{+}|\frac{d}{dt}|u_{+}\rangle&\sqrt{\frac{\Omega_-}{\Omega_+}}\langle v_{+}|\frac{d}{dt}|u_{-}\rangle{\rm e}^{{\rm i}\int_0^t(\Omega_+-\Omega_-)dt'}\\\sqrt{\frac{\Omega_+}{\Omega_-}}\langle v_{-}|\frac{d}{dt}|u_{+}\rangle{\rm e}^{-{\rm i}\int_0^t(\Omega_+-\Omega_-)dt'}&\langle v_{-}|\frac{d}{dt}|u_{-}\rangle\end{matrix}\right)\left(\begin{matrix}c_{+}\\c_{-}\end{matrix}\right)
    \label{eq:eqm_amplitude}.
\end{equation}
Within Eq. (\ref{eq:eqm_amplitude}) we have introduced the dual basis to the non--orthogonal vectors defined in Eq. (\ref{eq:ev})
\begin{align}
    \langle v_{+}|&=\frac{1}{\Omega_{+}^2-\Omega_{-}^2}\left(\frac{\Omega_{\mathcal{K}}^2-\Omega_{-}^2}{\bar{a}}{\rm e}^{\frac{\gamma t}{2}},1\right)\nonumber\\
    \langle v_{-}|&=\frac{1}{\Omega_{-}^2-\Omega_{+}^2}\left(\frac{\Omega_{\mathcal{K}}^2-\Omega_{+}^2}{\bar{a}}{\rm e}^{\frac{\gamma t}{2}},1\right)\label{eq:dual_vectors},
\end{align}
which obey the orthogonality relation $\langle v_{i}|u_{j}\rangle=\delta_{ij}$.  Using the expressions for the eigenvectors (\ref{eq:ev}), the inner products appearing in (\ref{eq:eqm_amplitude})---which are the analogue of the geometric phase~\cite{berry1984} for this non--Hermitian system---are equal to
\begin{equation}
    \langle v_{\pm}|\frac{d}{dt}|u_{\pm}\rangle=\pm\frac{\dot{\bar{a}}}{\bar{a}}\left(\frac{\Omega_{\mathcal{K}}^2-\Omega_\mp^2+\frac{1}{2}A\bar{a}}{\Omega_+^2-\Omega_-^2}\right)\mp\frac{\gamma}{2}\frac{\Omega_{\mathcal{K}}^2-\Omega_\mp^2}{\Omega_+^2-\Omega_-^2}+\frac{d}{dt}\ln\left(\sqrt{\Omega_+^2-\Omega_-^2}\right)
\end{equation}
and
\begin{equation}
    \langle v_{\mp}|\frac{d}{dt}|u_{\pm}\rangle=\mp\frac{\dot{\bar{a}}}{\bar{a}}\left(\frac{\Omega_{\mathcal{K}}^2-\Omega_\pm^2+\frac{1}{2}A\bar{a}}{\Omega_+^2-\Omega_-^2}\right)\pm\frac{\gamma}{2}\frac{\Omega_{\mathcal{K}}^2-\Omega_\pm^2}{\Omega_+^2-\Omega_-^2}-\frac{d}{dt}\ln\left(\sqrt{\Omega_+^2-\Omega_-^2}\right)
\end{equation}
Provided the frequencies $\Omega_{\pm}$ are real and the phase factors $\exp(\pm{\rm i}\int_0^t(\Omega_+-\Omega_-)dt')$ oscillate several times over the timescale where $\bar{a}$ changes significantly, the off diagonal elements of the matrix in (\ref{eq:eqm_amplitude}) can be neglected (corresponding to dropping an oscillatory integral) and the solutions for weak damping and slow speaker modulation are
\begin{equation}
    |\psi_+\rangle=\left(\begin{matrix}u_{\mathcal{K},+}\\v_+\end{matrix}\right)={\rm e}^{-\int_0^t\frac{\Omega_{\mathcal{K}}^2-\Omega_-^2+\frac{1}{2}A\bar{a}}{\Omega_+^2-\Omega_-^2}\frac{\dot{\bar{a}}}{\bar{a}}dt'}\frac{{\rm e}^{-{\rm i}\int_0^t\Omega_+dt'}{\rm e}^{-\frac{\gamma}{2}\int_0^t\frac{\Omega_+^2-\Omega_{\mathcal{K}}^2}{\Omega_+^2-\Omega_-^2}dt'}}{\sqrt{\Omega_+}\sqrt{\Omega_+^2-\Omega_-^2}}\left(\begin{matrix}\bar{a}(t)\\\Omega_+^2-\Omega_{\mathcal{K}}^2\end{matrix}\right)\label{eq:wkb1}
\end{equation}
and
\begin{equation}
    |\psi_-\rangle=\left(\begin{matrix}u_{\mathcal{K},-}\\v_-\end{matrix}\right)={\rm e}^{\int_0^t\frac{\Omega_{\mathcal{K}}^2-\Omega_+^2+\frac{1}{2}A\bar{a}}{\Omega_+^2-\Omega_-^2}\frac{\dot{\bar{a}}}{\bar{a}}dt'}\frac{{\rm e}^{-{\rm i}\int_0^t\Omega_-dt'}{\rm e}^{-\frac{\gamma}{2}\int_0^t\frac{\Omega_{\mathcal{K}}^2-\Omega_-^2}{\Omega_+^2-\Omega_-^2}dt'}}{\sqrt{\Omega_-}\sqrt{\Omega_+^2-\Omega_-^2}}\left(\begin{matrix}\bar{a}(t)\\\Omega_-^2-\Omega_{\mathcal{K}}^2\end{matrix}\right).\label{eq:wkb2}
\end{equation}
Solutions (\ref{eq:wkb1}--\ref{eq:wkb2}) become exact in the limit where there is no change in the speaker output $\bar{a}$, and $\gamma=0$.  The adiabatic approximation requires the system to be far from points where the acoustic frequencies change from real to complex.  This includes exceptional points where $\Omega_+=\Omega_-$, and points of vanishing frequency $\Omega_\pm=0$.  The failure of the approximation close to these points is evident in the divergence of the prefactors in Eqns. (\ref{eq:wkb1}--\ref{eq:wkb2}), just as for the ordinary WKB approximation~\cite{heading2013}.
%
%
\begin{figure}[h!]
    \includegraphics[width=1.0\textwidth]{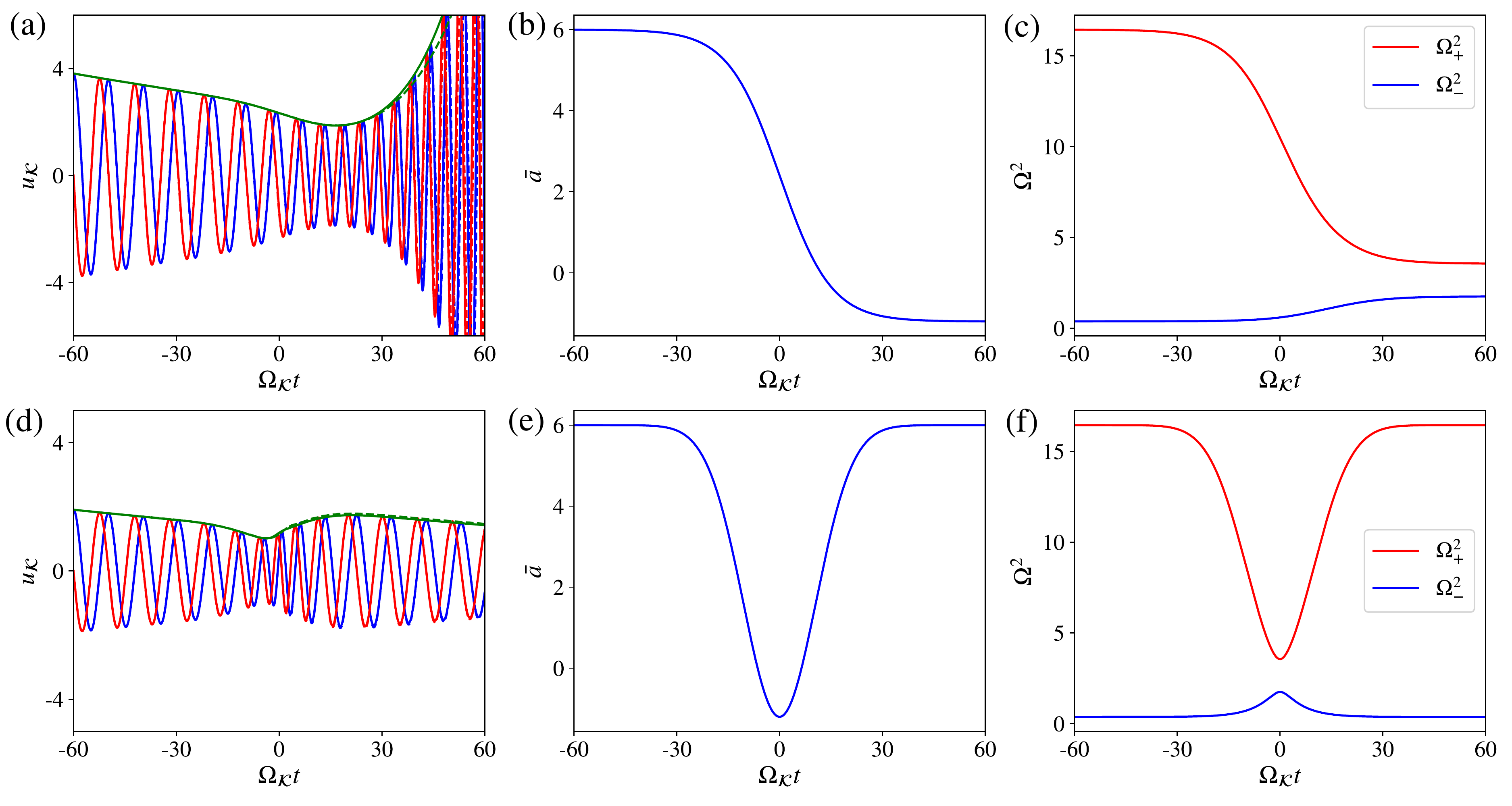}
    \caption{Adiabatic approximation (\ref{eq:wkb1}--\ref{eq:wkb2}) to the pressure field $u_{\mathcal{K}}$ with frequency $\Omega_-$. (a--c) Parameters as in Fig.~\ref{fig:dispersion}, with the speaker amplitude modulated from $\bar{a}=6$ to $\bar{a}=-1.2$, as a hyperbolic tangent $\tanh(t/T)$, with $\Omega_{\mathcal{K}}T=20$.  (a) Mode (blue: real part, red: imaginary part, green: absolute value) with frequency $\Omega_{-}$ smoothly changes from decay to amplification with time, despite approaching the exceptional point, as shown in panel (c). Numerical integration of (\ref{eq:two_oscillator_system}) shown as dashed lines. (d--f) Parameters as in Fig.~\ref{fig:dispersion}, illustrating the characteristic $1/\sqrt{\Omega_-}$ as the frequency is increased.\label{fig:wkbgood}}
\end{figure}
\par
For a constant coupling $\bar{a}$ the adiabatic solutions (\ref{eq:wkb1}--\ref{eq:wkb2}) oscillate with frequency $\Omega_{\pm}$ and decay or grow in time as $\exp(-\gamma_{\rm eff,\pm}t)$, where the effective damping constant $\gamma_{\rm eff,\pm}$ is given by
\begin{equation}
    \gamma_{\rm eff,\pm}=\pm\frac{\gamma}{2}\left(\frac{\Omega_{\pm}^2-\Omega_{\mathcal{K}}^2}{\Omega_{+}^2-\Omega_{-}^2}\right).\label{eq:gamma_eff}
\end{equation}
This effective damping constant is plotted in Fig.~\ref{fig:dispersion}b, showing that for a negative value of $\bar{a}A$ it can take either sign, while it remains positive for $\bar{a}A>0$.  Fig.~\ref{fig:dispersion}b also shows that as the speaker amplitude $\bar{a}$ approaches an exceptional point the effective damping diverges, indicating a failure of the approximation.  On the other hand, when $\bar{a}$ equals zero the acoustic mode is decoupled from the single board computer and has an effective damping constant of zero (red line in Fig.~\ref{fig:dispersion}).  Moving either side of this point, this mostly acoustic mode is either amplified or damped, depending on the sign of $\bar{a}$.  This is the expected behaviour of an acoustic wave coupled to a medium with Lorentzian response when the amplitude of the Lorentzian changes sign, switching from loss to gain. 
%
%
\par
In Fig.~\ref{fig:wkbgood}a,d we show a comparison between the adiabatic approximation (\ref{eq:wkb1}--\ref{eq:wkb2}) and a direct numerical integration of (\ref{eq:two_oscillator_system}), the two being almost indistinguishable.  In Fig.~\ref{fig:wkbgood}a the amplitude $\bar{a}$ changes sign in time and accordingly the $\Omega_{-}$ mode smoothly transitions from exponential decay to exponential growth, as discussed above. Fig.~\ref{fig:wkbgood}d also shows the accuracy of the adiabatic approximation when the amplitude $\bar{a}$ is modulated as a Gaussian, temporarily increasing $\Omega_-$, and decreasing the gap between the frequencies $\Omega_{\pm}$.  This increase in frequency causes the wave amplitude to be suppressed by the factor $1/\sqrt{\Omega_-}$ given in Eq. (\ref{eq:wkb2}).

At first sight these results are curious.  The adiabatic approximation apparently works well in a regime where $\Omega_-$ is `small' relative to the other frequencies, $\Omega_{+}$, $\omega_0$, and $\Omega_{\mathcal{K}}$.  Yet we stated earlier that the approximation ought to break down in this region.  The accuracy of the approximation for the $\Omega_-$ mode can be explained through examining the matrix on the right hand side of Eq. (\ref{eq:eqm_amplitude}).  To obtain (\ref{eq:wkb1}--\ref{eq:wkb2}) we dropped the off diagonal terms which are proportional to $\sqrt{\Omega_{+}/\Omega_{-}}$ and $\sqrt{\Omega_{-}/\Omega_{+}}$.  When $\sqrt{\Omega_{-}/\Omega_{+}}\ll1$ only the top right diagonal element of this matrix can be neglected.  The resulting triangular matrix has only one eigenvector, which is the mode with frequency $\Omega_-$.  Therefore it is the other ($\Omega_+$) mode that is a poor approximation in this regime.
%
%
\begin{figure}[h!]
    \includegraphics[width=\textwidth]{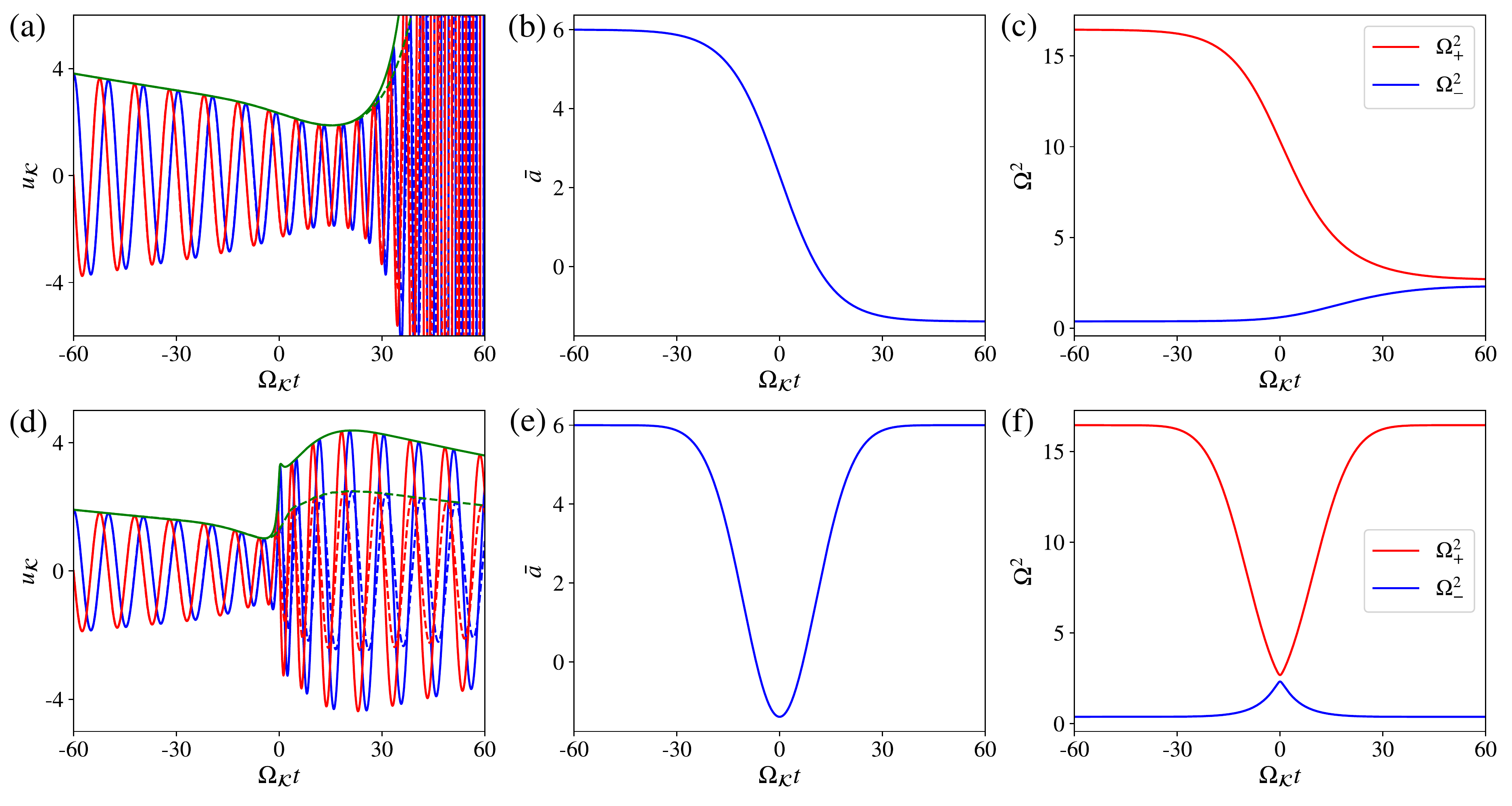}
    \caption{Speaker modulation $\bar{a}(t)$ where the adiabatic approximation (\ref{eq:wkb1}--\ref{eq:wkb2}) fails.  Parameters and quantities plotted as in Fig.~\ref{fig:wkbgood}, but for modulation of the coupling between $\bar{a}=6$ and $\bar{a}=-1.38$, the final value being very close to the exceptional point where $\Omega_+$ and $\Omega_-$ are degenerate. Fig.~\ref{fig:dispersion}b shows that the effective damping in the adiabatic approximation becomes negative and divergent at the exceptional point.  Accordingly panels (a) and (d) show the adiabatic approximation (solid) overestimates the amplification of the wave (numerical solution of (\ref{eq:two_oscillator_system}), shown as dashed line). \label{fig:wkbbad}}
\end{figure}
\par
Conversely Fig.~\ref{fig:wkbbad} compares adiabatic and numerical solutions when the coupling $\bar{a}$ is modulated to a value that is $99\%$ of $\bar{a}_{\rm ex}$.  While numerical and adiabatic solutions agree well before the modulation, they differ significantly as the exceptional point is approached.  For instance the adiabatic approximation overestimates the amplification of the wave in both Fig.~\ref{fig:wkbbad}a and Fig.~\ref{fig:wkbbad}d.  This is closely related to the well known failure of the adiabatic approximation in non--Hermitian systems, close to an exceptional point~\cite{heading2013,nenciut1992,uzdin2011,berry2011a,berry2011b}.  In time varying systems such as our array of digital meta--atoms, slow parameter change does not necessarily correspond to an adiabatic transition between the instantaneous eigenstates.
%
%
\section{Non--adiabatic modulation}
\par
In the opposite limit, we can consider an abrupt change in the speaker output amplitude at $t=0$, between two constant values, $\bar{a}_1$ and $\bar{a}_2$.  This abrupt change will lead to the temporal analogue of reflection, as discussed in e.g.~\cite{mendonca2002} where the positive and negative frequency solutions to (\ref{eq:two_oscillator_system}) will become mixed after the abrupt change.  Using Eqns. (\ref{eq:wkb1}--\ref{eq:wkb2}), for $t>0$ we take our field to be a sum over the four approximate solutions to (\ref{eq:two_oscillator_system})
\begin{equation}
    |\psi(t>0)\rangle=\alpha_a|\psi_a\rangle+\alpha_b|\psi_b\rangle+\alpha_c|\psi_c\rangle+\alpha_d|\psi_d\rangle
\end{equation}
where the four modes $|\psi_{a,b,c,d}\rangle$ are given by
\begin{equation}
    |\psi_{a,b}\rangle=\left(\begin{matrix}\bar{a}_{2}\\\Omega_{2,+}^2-\Omega_{\mathcal{K}}^2\end{matrix}\right){\rm e}^{-{\rm i}\omega_{a,b} t},\quad|\psi_{c,d}\rangle=\left(\begin{matrix}\bar{a}_{2}\\\Omega_{2,-}^2-\Omega_{\mathcal{K}}^2\end{matrix}\right){\rm e}^{-{\rm i}\omega_{c,d}t}\label{eq:sol_abcd}
\end{equation}
where the complex frequencies are given by $\omega_{{}^{a}_{b}}=\pm\Omega_{2,+}-({\rm i}\gamma/2)(\Omega_{2,+}^2-\Omega_{\mathcal{K}}^2)(\Omega_{2,+}^2-\Omega_{2,-}^2)^{-1}$ and $\omega_{{}^{c}_{d}}=\pm\Omega_{2,-}+({\rm i}\gamma/2)(\Omega_{2,-}^2-\Omega_{\mathcal{K}}^2)(\Omega_{2,+}^2-\Omega_{2,-}^2)^{-1}$.  Before $t=0$ the field is taken as a single mode, corresponding to the $\Omega_{-}$ root of (\ref{eq:Wpm})
\begin{equation}
    |\psi(t<0)\rangle=\left(\begin{matrix}\bar{a}_1\\\Omega_{1,-}^2-\Omega_{\mathcal{K}}^2\end{matrix}\right){\rm e}^{-{\rm i}\omega_0 t}
\end{equation}
where $\omega_0=\Omega_{1,-}+({\rm i}\gamma/2)(\Omega_{1,-}^2-\Omega_{\mathcal{K}}^2)(\Omega_{1,+}^2-\Omega_{1,-}^2)^{-1}$.  Applying the continuity of the fields $u_{\mathcal{K}}$ and $v$ and their time derivatives at $t=0$ we can determine the four unknowns $\alpha_{a,b,c,d}$ and thereby find the temporal analogue of the reflection coefficients, which are
\begin{align}
    \alpha_{a}&=\frac{(\Omega_{1,-}^2-\Omega_{\mathcal{K}}^2)-\left(\frac{\bar{a}_1}{\bar{a}_2}\right)(\Omega_{2,-}^2-\Omega_{\mathcal{K}}^2)}{\Omega_{2,+}^2-\Omega_{2,-}^2}\frac{\omega_b-\omega_0}{\omega_b-\omega_a}\nonumber\\
    \alpha_{b}&=\frac{(\Omega_{1,-}^2-\Omega_{\mathcal{K}}^2)-\left(\frac{\bar{a}_1}{\bar{a}_2}\right)(\Omega_{2,-}^2-\Omega_{\mathcal{K}}^2)}{\Omega_{2,+}^2-\Omega_{2,-}^2}\frac{\omega_0-\omega_a}{\omega_b-\omega_a}\label{eq:alpha_ab}
\end{align}
and
\begin{align}
    \alpha_{c}&=\frac{(\Omega_{1,-}^2-\Omega_{\mathcal{K}}^2)-\left(\frac{\bar{a}_1}{\bar{a}_2}\right)\left(\Omega_{2,+}^2-\Omega_{\mathcal{K}}^2\right)}{\Omega_{2,+}^2-\Omega_{2,-}^2}\frac{\omega_0-\omega_d}{\omega_d-\omega_c}\nonumber\\
    \alpha_{d}&=\frac{\left(\Omega_{1,-}^2-\Omega_{\mathcal{K}}^2\right)-\left(\frac{\bar{a}_1}{\bar{a}_2}\right)\left(\Omega_{2,+}^2-\Omega_{\mathcal{K}}^2\right)}{\Omega_{2,+}^2-\Omega_{2,-}^2}\frac{\omega_c-\omega_0}{\omega_d-\omega_c}.\label{eq:alpha_cd}
\end{align}
%
%
\begin{figure}[h!]
    \includegraphics[width=\textwidth]{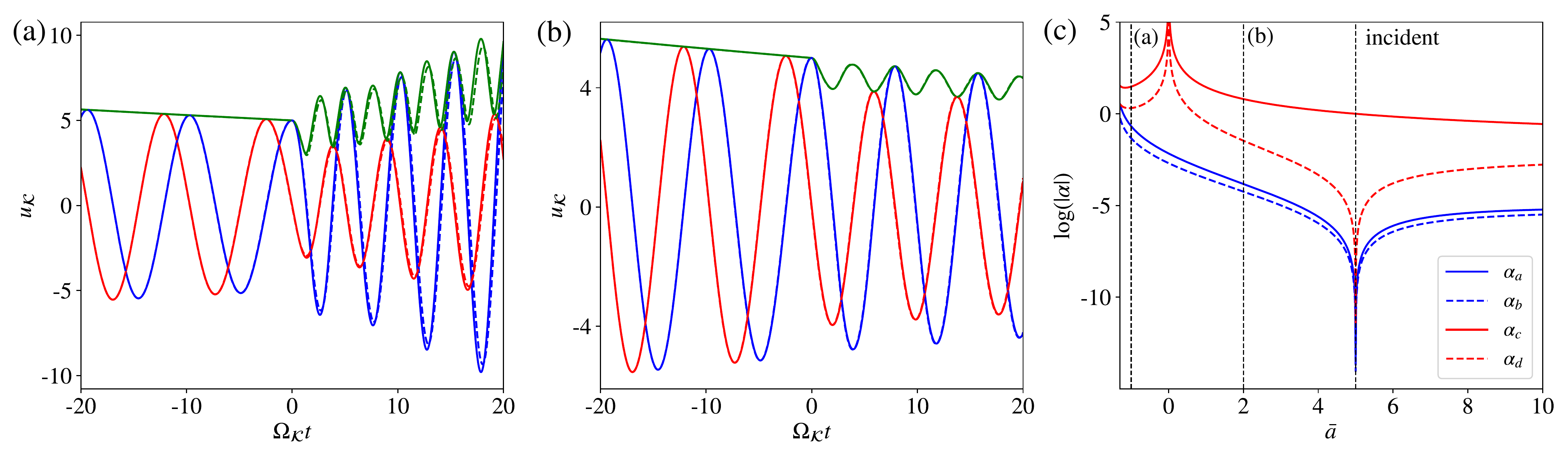}
    \caption{Reflection and mode conversion after a sudden change in the speaker output $\bar{a}$.  Parameters as in Fig.~\ref{fig:dispersion}.  Panel (c) shows the logarithm of the four mode amplitudes (\ref{eq:alpha_ab}--\ref{eq:alpha_cd}) after the modulation, where the initial speaker output is $\bar{a}=5$, and the final amplitude is varied.  The divergence of $\alpha_{a,b}$ at $\bar{a}=0$ is due to the vanishing of the vectors $|\psi_{a,b}\rangle$ defined in (\ref{eq:sol_abcd}), and does not represent any divergent physical quantity.  Panels (a) and (b) show the real (blue), imaginary (red), and absolute value (green) of the pressure field for the two final values of $\bar{a}$ indicated in (c).  Results of numerical integration of (\ref{eq:two_oscillator_system}) are shown as dashed lines.\label{fig:reflection}}
\end{figure}
The four distinct modes arise from the dispersion in this system and accordingly the `reflection' coefficients involve both the conventional mixing of forwards and backwards propagating waves (i.e. amplitudes corresponding to the $\omega_{{}^{a}_{b}}$ or $\omega_{{}^{c}_{d}}$ modes alone), and mode conversion (e.g. mixing between $\omega_{{}^{a}_{b}}$ and $\omega_{{}^{c}_{d}}$ modes).

The distinction between reflection and mode conversion is illustrated in Figs.~\ref{fig:reflection}b and c, where far from the exceptional point the dominant amplitudes after the modulation are of the same type (shown in red) as the incident wave (i.e. the modes are mostly the $\Omega_{-}$ root of the dispersion relation, before and after the modulation).  Meanwhile, close to the exceptional point (see Figs.~\ref{fig:reflection}a and c), there is much stronger coupling between the different roots of the dispersion relation.  As shown in Fig.~\ref{fig:reflection}c, the $\Omega_{2,-}$ and $\Omega_{2,+}$ waves (red and blue, respectively) are excited in similar proportions by changing $\bar{a}$ to a value that is close to the exceptional point value $\bar{a}_{\rm ex}$, defined in (\ref{eq:aex}).  Although we have derived the mode amplitudes (\ref{eq:alpha_ab}--\ref{eq:alpha_cd}) using the adiabatic approximation for $t<0$ and $t>0$, Fig.~\ref{fig:reflection} shows excellent agreement with a direct numerical integration of the equations of motion.  The agreement cannot be expected to be as good for incidence of the $\Omega_{1,+}$ mode, for the reason explained in the previous section.

\section{Summary and conclusions}
\par
We have developed a model of acoustic mode propagation through an array of digital meta-atoms, active elements that can be programmed to give an almost arbitrary dispersive, time varying response.  Taking the system as an array of coupled point--like sources and detectors---where the source output is varied in time---we find the effective medium limit is equivalent to a dispersive medium, where the effective susceptibility is multiplied by a time dependent modulation, $\bar{a}(t)$.  Re--writing the dispersive system as a pair of coupled harmonic oscillators, we derived an approximate solution to the equations of motion for a fixed wave number $\mathcal{K}$, valid for the lowest frequency mode when there is slow modulation of $\bar{a}$ and weak damping, $\gamma$.  This solution was verified numerically, where the modulation of the output $\bar{a}$ was chosen to smoothly modify the acoustic mode from one that is damped, to one that grows in time.  The adiabatic solution also captures the change in wave amplitude when there is a significant change in the mode frequency.

In general the response of the digital meta--atom array is both non--Hermitian and time varying, and accordingly exhibits exceptional points for some choices of output amplitude.  We find our adiabatic solution breaks down close to these exceptional points, where there is a strong coupling between the modes.  This shows that the intuition that mode conversion requires the system parameters to be varied on the time--scale of the wave period does not hold in non--Hermitian systems, and may be deliberately broken in such active meta--materials.  This may be a useful feature in e.g. time varying electromagnetic media when the wave period is shorter than the possible meta--atom response time.

We also examined the evolution of the wave in the opposite limit, where the speaker amplitudes $\bar{a}$ are changed abruptly in time.  Using our adiabatic solution either side of the abrupt change and applying continuity at the interface, we derived formulae for the final mode amplitudes, verifying the formulae numerically.  These `reflection coefficients' describe both the generation of negative frequency waves (change in propagation direction), and the conversion between different modes of the system.  This is analogous to e.g. a spatial interface between two different multi--mode waveguides, where the interface causes both reflection between like modes, and conversion between different waveguide modes.

While the change in propagation direction is already well known in the non--dispersive approximation to time varying media, the dispersive response controls the \emph{number} of modes supported at a fixed wave number $\mathcal{K}$.  As a digital meta--atom can be programmed at will, there is the potential to manipulate this multi--mode conversion phenomenon via such time varying active media.
%
%
\acknowledgements
SARH acknowledges funding from the Royal Society and TATA (RPG-2016-186) as well as useful comments from Gianluca Memoli, and Bryn Davies.
%
%
\bibliography{refs}
\end{document}